# Research Portfolio Analysis and Topic Prominence

Richard Klavans[*] and Kevin W. Boyack[**]

[*] *rklavans@mapofscience.com*
SciTech Strategies, Inc., Wayne, PA 19087 (USA)

[**] *kboyack@mapofscience.com*
SciTech Strategies, Inc., Albuquerque, NM 87122 (USA)

**Abstract**

Stakeholders in the science system need to decide where to place their bets. Example questions include: Which areas of research should get more funding? Who should we hire? Which projects should we abandon and which new projects should we start? Making informed choices requires knowledge about these research options. Unfortunately, to date research portfolio options have not been defined in a consistent, transparent and relevant manner. Furthermore, we don't know how to define demand for these options. In this article, we address the issues of consistency, transparency, relevance and demand by using a model of science consisting of 91,726 topics (or research options) that contain over 58 million documents. We present a new indicator of topic prominence – a measure of visibility, momentum and, ultimately, demand. We assign over $203 billion of project-level funding data from STAR METRICS® to individual topics in science, and show that the indicator of topic prominence, explains over one-third of the variance in current (or future) funding by topic. We also show that highly prominent topics receive far more funding per researcher than topics that are not prominent. Implications of these results for research planning and portfolio analysis by institutions and researchers are emphasized.

## 1 Introduction

Research portfolio analysis should be a key activity for all stakeholders in the current science system. Funding bodies allocate resources among topics, administrators choose which researchers to hire and which projects to support internally, while researchers (for the most part) choose the topics they want to work on (Fisher, 2005; Foster, Rzhetsky, & Evans, 2015; Zuckerman, 1978). The notions of research portfolios and portfolio analysis, once largely confined to the corporate R&D world, are now being increasingly considered in academic and agency settings (Wallace & Rafols, 2015).

Portfolio-related choices are, however, difficult to make in the science system because the potential choices themselves are often not well defined or understood. In the industrial world, research portfolio choices are typically governed by perceptions of (long-term) supply and demand. We suggest that the concepts of supply and demand can also provide a useful framework for research portfolio analysis in the science system. However, to use these terms we must take care to define them properly as they can be defined in different ways.

For instance, Sarewitz & Pielke (2007) define supply and demand at a high level in terms of the interplay between scientific results and their providers (supply) and specific societal goals (demand). At a more detailed level, Sarewitz & Pielke define demand in terms of the information used by a wide variety of stakeholders to address a broad set of challenges (p. 12). Dalrymple (2017) defines supply as the output of the public research process and demand as reflecting the interests of the users. These interests – the inputs to demand – can vary. The demand for science to address social and health needs is prevalent, and can be represented by metrics such as disease burden (Atanassova, 2016). Some science, particularly that invested in by industry, responds to economic motives (Klavans & Boyack, 2017a). These diverse demands are important, and they (and their advocates) obviously play a role in priority setting within governments, agencies, and other funding bodies. It is also true that these types of demand are very difficult to measure. Rather than considering each type of demand separately, we simply note that the ultimate result of these varying demands is that research priorities are set and funding is accordingly made available to address these priorities. Accordingly, we make explicit this assumption that funding amounts represent an aggregate (though undoubtedly crude and incomplete) measure of demand for each priority or topic in science. Thus, our definition of demand differs from those of Sarewitz & Pielke and Dalrymple in that it assumes that interests and goals are codified in a fiscal sense. The scientific topics themselves – the outputs of science mentioned by Sarewitz & Pielke and Dalrymple – represent the scientific supply.

Other frameworks are possible. For instance, another common framework uses the language of investments and outputs (Lewis, 2003; Wang & Shapira, 2015). Funders invest in research, while researchers create outputs (articles) that may influence social outcomes. While this framework is valid (it is commonly used to analyse securities as well as grant investments), we prefer the more common definition of portfolio analysis associated with commerce (http://www.businessdictionary.com/definition/portfolio-analysis.html). Research is viewed as a product (publications represent supply) while funders purchase that research (grants represent demand).

With this framework in place, the gap that remains is the lack of a detailed model of scientific topics that represents supply and to which demand can be linked. Accordingly, this study introduces a topic-level model of the scientific literature. We define a topic as a collection of documents with a common focused intellectual interest, such as the work on a specific research problem. We then go further by developing a new indicator of topic prominence, and show that this indicator is a good predictor of current and future funding at the topic level, and is thus an indicator of demand. Unlike previous bibliometric studies, where the emphasis has only been on identifying emerging topics (Small, Boyack, & Klavans, 2014) or research fronts (Clarivate, 2016), our goal is to look at the entire portfolio of choices. As such, this is the first large-scale test of a highly detailed portfolio model of research.

The article is organized as follows. The background section starts by describing the theory and practice underlying the identification and evaluation of all possible research topics in the scientific literature. Critical to this discussion are the issues of coverage, granularity, accuracy and stability. We also briefly discuss potential indicators of topic impact and relevance. Additional sections then detail the four main contributions of the article: 1) creation and description of the topic-level model of science, 2) formulation of an indicator of topic prominence, 3) the assignment of project-level grant data to individual topics and 4) the use of the prominence indicator to explain and predict topic funding levels. Each of these sections combines methods and results. The article closes with a discussion of weaknesses and limitations of the study, and with implications for research planning by funding agencies, research institutions and individual researchers.

## 2 Background

The structure of science, as a whole or by parts, can be represented in many different ways including journal subject categories, controlled vocabularies such as MeSH, or clusters based on citation or textual characteristics. The funding of science has been linked to its structure, but this has been done only at very high levels. For instance, R&D expenditures have been reported at the level of eight S&E fields (National Science Board, 2001), by agency or sub-agency, and by disease category (Atanassova, 2016; Bornmann & Daniel, 2007). We are unaware of any studies that report funding at much more detailed levels, perhaps because more detailed classification systems are not commonly available.

Creation of a detailed topic-level model of science requires many design choices, the most important of which are related to coverage, granularity, accuracy and stability. Accordingly, this section focuses on these choices and on the theoretical and historical bases that give us reasonable guidance as to make these choices. We note, however, that we are operating within the context of using citation data to create models. Models created using textual characteristics or controlled vocabularies (such as MeSH) will have different properties. Such models may be appealing to different communities. For instance, disease categories using MeSH terms have been used by policy makers interested in comparing funding with disease burden for a defined set of diseases (Evans, Shim, & Ioannidis, 2014).

### 2.1 Topic Coverage

Our goal is to identify all topics in science for the use of multiple stakeholders. Thus, the ideal would be to have access to all literature on all topics, and to then have a way to partition that literature into topics. This ideal, however, may not currently be reachable since no single database covers all of the scientific literature. In addition, there are differences of opinion over whether a full database is necessary to accurately identify topics or if smaller datasets using journals or keyword searches are adequate.

Although not definitive, there is a literature that addresses these issues and provides some guidance. Regarding database coverage, there are two large citation databases (Scopus and the Web of Science (WoS)) that cover a significant fraction of the scientific literature. However, it is also well known that coverage within these databases varies by field. Each is known to have very good coverage in medicine, but lower coverage in the social and computer sciences (Boyack & Klavans, 2014b; Hicks, 2004; Moed, 2005). This imbalance may influence the identification of topics. There are also cases where the two large citation databases are potentially inadequate for the identification of topics. For instance, Rafols et al. (2015) showed that the CAB Abstracts (CABI) database, which focuses on environmental and agricultural literature, contains well over twice as many documents about rice as do the citation databases. In particular, CABI had much better coverage on the topics of productivity, plant nutrition, plant characteristics and plant protection. Use of either Scopus or WoS may miss some specific topics altogether. Despite their deficiencies, given the need to identify all topics in science, the two large citation databases remain the best choices.

We turn now to the question of whether entire databases should be used to identify topics, or if smaller datasets can be used. In other words, can local mapping be used to accurately identify topics, or is global mapping a far superior technique. Several of our recent studies address this issue. In the first, we compared topics related to information science from a global model of science to the 50 topics identified by Chen et al. (2010) using co-citation analysis of papers from 12 information science journals (Klavans & Boyack, 2011). We found that while the local analysis did a good job of identifying core topics in information science, it under-specified the topics that were dominated by literature from another field (e.g., computer science). A similar example is found in Boyack (2017), where the topics associated with articles in a set of 59 journals in astronomy, which is considered to be an insular discipline, were found to be embedded in a network that contained nearly as many papers outside those journals. This study also found that articles where the fraction of within-local set links (as compared to the global links) was small were likely to be part of topics that were not based in astronomy, despite the fact that the articles were in astronomy journals.

A third study showed that the same type of behavior occurs when using a dataset based on keyword searches rather than a journal set (Boyack, Klavans, Small, & Ungar, 2014). This study showed that many topics now associated with 'graphene' were involved in other research activities (such as research on graphite or carbon nanotubes) that were later replaced by research on graphene once graphene became available as a research material.

In summary, the studies referenced above have shown that a local model typically excludes many papers that are highly linked to either journal or keyword based datasets (lower recall) and includes papers that are not well linked within the set and thus belong elsewhere (lower precision). Global models, by definition, include much more of the relevant content. Thus, local models have both lower recall and lower precision than is desirable for the accurate identification of topics. In local models, papers with low within-local set linkage are placed in the core topics rather than multidisciplinary topics due to the exclusion of global material. Indeed, topics identified using subsets of the literature may completely miss interdisciplinary topics or topics at the boundary of the field of interest. It is generally accepted that multiple iterations of query expansion and filtering can increase precision and recall, yet it is unclear how well this really does at bounding a field because most studies of query expansion have no ground truth against which to compare.

One might think that, given the problem of coverage, scientometricians would lean towards identifying topics from full databases. However, this has not happened for pragmatic reasons. Very few researchers have access to the entire WoS or Scopus database. Researchers can, however, download small portions of these databases via institutional licenses. It is for this reason that studies using small datasets far outnumber studies using an entire database. These small datasets may be sufficiently representative to make conclusions about the major structural elements. However, topics that tend to be interdisciplinary cannot be accurately identified using small datasets.

## 2.2 Number of Topics

The granularity of a model of science – the number and size of research topics – is also important to portfolio analysis. Most evaluation using normalized citation-based indicators (Waltman, 2016) is currently done using approximately 250 journal subject categories. While this is likely an appropriate level for some types of analysis, it is not nearly detailed enough to enable researchers to choose between topics. We have found that researchers and program officers at funding agencies can easily differentiate (and tell stories about) roughly 100,000 topics in science (Boyack, et al., 2014).

An order of magnitude approach that incorporates both theory and measurements of current science also leads to an estimate of around 100,000 topics. Early work by science historians and sociologists of science (Crane, 1972; Kuhn, 1970; Mullins, 1973) discussed research communities that, once mature, would be addressing a single scientific specialty (or topic) and would be comprised of about 100 researchers. Some communities will be larger and some will be smaller. If we assume that each topic has on average 100 researchers, then the total number of current authors gives us a rough sense of the number of communities (or topics) that exist in the world. A recent effort to disambiguate author names over the entire Web of Science found 6.7 million authors with an h-index of greater than one (Schulz, Mazloumian, Petersen, Penner, & Helbing, 2014). A query of Scopus data (as of 2015) shows that there are over 10 million unique author IDs associated with at least one publication over the most recent four years. Thus, it seems reasonable that with around 10 million currently publishing authors, there may be on the order of 100,000 topics in science today.

In addition, we have compared models with 100,000 and 10,000 topics and have found that sets of key papers in topics from the former model are focused on the same narrow research problem. In contrast, sets of key papers from the latter model appear to address different, although related, research problems. We also find it easier to describe topics (based on perusal of their representative keywords and key papers) in the model with 100,000 clusters than those in the model with 10,000 clusters. This is admittedly anecdotal, but does represent our experience with models at different levels. Given that our focus is on portfolio analysis that represents choices between research problems, we choose to create models with roughly 100,000 clusters.

## 2.3 Topic Accuracy

The accuracy of topics is critical for portfolio analysis. One school of thought is that different approaches give different results that are each meaningful and that the accuracy of different approaches cannot really be established (Gläser, Glänzel, & Scharnhorst, 2017). This is a reasonable stance to take if one wants to learn about different approaches, their tendencies, and how they might differently portray topics. However, real-life decisions to the types of questions that are asked by stakeholders in the science system require a more definitive answer. Stakeholders want to base decisions on the most accurate depiction of topics available.

The accuracy of a particular model is best established by referring to a gold standard that provides a fair basis of comparison. Conceptually, it seems reasonable that this basis should be independent of the methods being compared (Waltman, Boyack, Colavizza, & Van Eck, 2017).

Fortunately, there are several recent studies that compare the accuracy of different approaches to identifying topics. In a large study comparing citation-based and text-based approaches, Boyack and colleagues (Boyack & Klavans, 2010; Boyack, et al., 2011) used grant-article linkages to compare different approaches under the assumption that if several articles cite the same grant they are very likely to be on the same topic, and should thus be concentrated (rather than dispersed) within the topic structure. Accordingly, a concentration index was used as the measurement technique, resulting in the finding that the best citation-based (bibliographic coupling) and text-based (PubMed related articles) gave roughly similarly accurate topics, and that a citation-text hybrid was slightly better than either.

A much larger study was undertaken more recently. Klavans & Boyack (2017b) compared detailed topic-level structures of up to 45 million documents created using direct citation (including the model from Section 3), bibliographic coupling, and co-citation methods. In this case, the lists of references in articles with at least 100 references were used as the gold standard under the assumption that researchers are authorities who understand and portray the topical structure of science in their papers. Clearly, this basis of comparison is not independent from the methods being measured because all of the methods were citation-based, and the concentration of citations is used as the basis of comparison. However, this method did have one strong advantage over other methods. It included the references to non-source items as part of the measurement, thus overcoming the effect of missing documents. 45% of the references in documents indexed in Scopus from 1996-2014 are to non-source items. By including these references in the basis of comparison, we nearly doubled the signal with which to compare relative accuracies. In this study, topics identified using extended direct citation (which includes cited non-source items) were found to be slightly more accurate than those identified using bibliographic coupling, and much more accurate than those identified using co-citation.

Finally, Waltman et al. (2017) recently introduced a principled approach to comparing methods, and used a text-based similarity to compare results from several citation-based approaches. Principled in this case means that the basis of comparison was independent from the approaches being compared. Using a set of nearly 273,000 documents in physics, they found that extended direct citation (which includes non-source documents in the clustering), bibliographic coupling, and a method combining direct citation, co-citation and bibliographic coupling all performed equally well. Each of these methods was found to be much more accurate than co-citation.

Despite the work done to establish the accuracy of topics, we acknowledge that the idea of accuracy in clustering is still not completely understood. It is likely that new methods for identifying topics and for measuring accuracy will be developed in the future that are better than those in use today. Moreover, is also true that each methodology entrains certain tendencies that may bias accuracy measurements. In the interest of openness, we list here some experience-based assumptions and tendencies associated with our methodology. We assume that citation signals are less ambiguous than textual signals. We assume that interdisciplinary topics cannot be adequately represented using small datasets. Related to the use of direct citation, topics are more associated with historical development than with semantic cohesion. Topics tend to have sizes that are reflective of the relative citation densities of their dominant fields. We do not know all of the ramifications of our design choices. We do know, however, that users resonate with topics and feel they are useful, which is a very positive, though entirely qualitative, way to describe the robustness of the methodology.

Overall, the evaluation of topic accuracy is still an open issue. We currently advocate using review papers as a basis of comparison in that we consider them to be expert judgments. There are two recent studies using different bases of comparison that both show that extended direct citation is at least as accurate as any other citation-based approach for identifying topics. Another benefit of using extended direct citation is that it is relatively simple and efficient because second-order links do not need to be calculated. Calculation of second-order links (i.e., bibliographic coupling or co-citation links) becomes prohibitively expensive for very large databases due to computational speed and memory limitations.

**2.4 Topic Stability**

Topic stability is another issue that needs to be considered. It has been our experience that stakeholders require topics that are relatively stable over time to monitor their portfolios year over year in a way that meets their needs. Planning and evaluation are far more difficult if topics are unstable. Users want to see trends, which are only observable if one has a stable set of topics. This does not mean that new topics cannot be born. Rather, it means that topics should not re-align themselves with the influx of publications from the most recent year. As such, a topic classification system that is highly stable is desirable.

Topic identification based on linking annual models using overlaps in clusters produces results that suggest science is unstable. Transience in year-to-year distributions was shown early on by Price & Gürsey (1975), who found that typically less than half of the authors publishing a paper one year also published one the next year. Small (1976) found the same thing for cited references. Only half of articles cited above a particular threshold were cited at that same level the next year. In linking topics from annual co-citation models, Boyack & Klavans (2014a) found that about 44% of research fronts do not continue (i.e., they die) in the next time period. If linking thresholds were set higher, no threads formed. If thresholds set were lower, topics very quickly coalesced into a giant component. Small (2009) observed similar behavior when linking individual documents into co-citation clusters.

In contrast to co-citation clusters, topics identified using direct citation and time windows of at least 10 years, while they have a great deal of dynamic variation, are also very stable in that most topics last for many years with relatively low birth and death rates (Small, et al., 2014). Most research grants last several years at a minimum, with a single investigator working on the grant topic continually over that time period. This suggests that topics should persist for at least that long. The combination of stability and dynamics associated with direct citation clusters suggest that they are appropriate for the purposes of portfolio analysis.

**2.5 Topic Characteristics for Portfolio Analysis**

The two characteristics that are commonly used in (commercial) portfolio analysis are market share and market growth. Relative publication share can be used to characterize market share. The following discussion focuses on the more difficult issue of how to determine if a topic is pre-emergent, growing or mature. Two indicators that have been applied at the topic level to indicate emergence or newness are emergence potential (Small, et al., 2014) and mean topic year (Clarivate, 2016). These two indicators both seem to work reasonably well to identify emerging topics, but they are both retrospective with a time lag of a couple of years. An indicator that can predict if a topic will grow or decline in the future would be very useful for planning and portfolio analysis. We detail the development of such an indicator later on, and suggest that any such indicator, if valid, should be able to predict the future allocation of grant funds. We also note that measures of societal importance would be extremely important to portfolio analysis. Although this paper does not focus on this type of metric, we feel development of such metrics is crucial to future advances.

## 3 Model of Research Topics

In this section, we describe the SciTech Strategies (STS) model of science along with the method by which it was created. This model is the "DC5" model whose topics were recently shown to be very consistent with the referencing patterns of experts (Klavans & Boyack, 2017b) and is consistent with design choices presented in the background section. For instance, we choose to use a complete citation database, and to partition the set of documents into around 100,000 topics. We choose to do this using an extended direct citation method that includes non-source items to increase the accuracy and stability of the topics. These choices are aimed at creating accurate topics at a granularity that is useful for stakeholders and their needs related to portfolio analysis and planning. We note that other choices could be made and justified for a different set of requirements. For instance, the Centre for Science and Technology Studies (CWTS) at Leiden University uses a similar approach, but chooses to create roughly 4000 categories (Ruiz-Castillo & Waltman, 2015) for use in normalizing indicators for the Leiden Ranking.

The original STS model of science consists of 48,398,815 documents (of all types) from Scopus. Of these, 24,615,844 documents are indexed source documents from Scopus 1996-2012, while the remaining 23,782,971 are non-source documents that were each cited at least twice by the set of source documents. Including non-source documents extends the coverage of the model to include important science not indexed by Scopus, including many books. In particular, the social sciences are heavily augmented when non-source documents are included (Boyack & Klavans, 2014b). The 48.4 million documents are connected by 582 million direct citation links, which were used to create the model. Clustering was done using the smart local moving (SLM) algorithm from CWTS (Waltman & van Eck, 2013) that has recently been shown to be among the most accurate available (Emmons, Kobourov, Gallant, & Börner, 2016). The similarity value ($a_{ij}$) between each pair of papers $i$ and $j$ was set to $1/k$ where $k$ is the number of edges (both citing and cited) for paper $j$ (Waltman & van Eck, 2012). Symmetry was not assumed; $a_{ij}$ and $a_{ji}$ are typically different since the numbers of edges for papers $i$ and $j$ are rarely the same. The SLM algorithm was run using the file of document-document similarities as input and a resolution of $3 \times 10^{-5}$. A cluster solution was obtained with 91,726 clusters above that each contained at least 50 articles. Each cluster represents a topic, and is comprised of the papers on that topic and the community of researchers working on that topic. 134,066 (0.28%) of the documents ended up in clusters with fewer than 50 documents. These clusters are small and disconnected from the rest of the graph, and thus are not considered further.

Figure 1 shows a map of the 91,726 topics created to provide a visual depiction of the structure of science. It was created using the following process: 1) similarity between pairs of topics was calculated from the titles and abstracts of the documents in each topic using the BM25 text similarity measure (Sparck Jones, Walker, & Robertson, 2000), 2) the resulting similarity list was filtered to keep only the top-n (between 5 and 15) similarities per topic, and 3) a layout of the topics was generated using the DrL algorithm (Martin, Brown, Klavans, & Boyack, 2011), which gives each topic an x,y position based on the similarity graph. One might wonder why an additional direct citation step is not used to create the visual map. This could certainly be done, but we have found that using text creates a more accurate and visually appealing map than using a citation-based measure for this step (Boyack & Klavans, 2014a). Each of the 91,726 topics in the map has been designated as belonging primarily to one of twelve high-level fields, and is colored accordingly. These designations were made by assigning papers to fields using the UCSD journal-to-field assignments (Börner, et al., 2012).

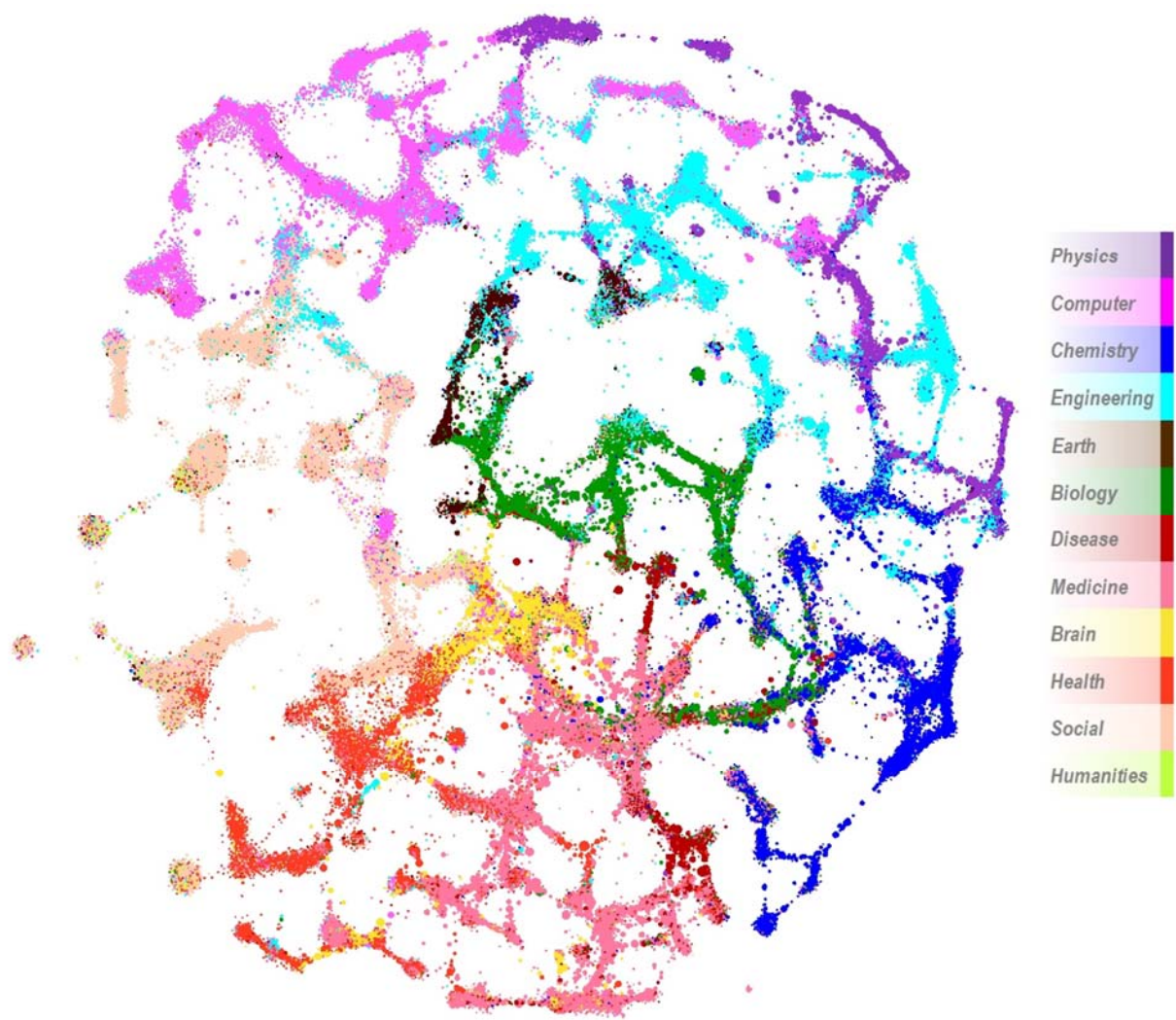

**Figure 1. Visual map of the STS model of science. Each dot represents a topic.**

At a high level, the field structure shown in Figure 1 is very similar to that of many other global science maps, including the consensus map of science (Klavans & Boyack, 2009). Physics, chemistry and engineering are highly related, and are adjacent to each other. The medical areas (disease, medicine, health sciences, brain sciences) are also adjacent to each other. Biology is adjacent to chemistry and medicine, earth sciences is primarily adjacent to engineering, and social sciences are adjacent to health sciences, while computer science lies between physics (which includes mathematics) and the social sciences.

This model has a number of features that attest to its robustness and that make it useful for a variety of applications. In terms of robustness, nearly 90% of the source documents in the map are found in the cluster to which they have the greatest number of links (combined citing and cited), and on average 58% of their links are to that dominant cluster. There is also a relatively high correlation between text and links in the model. Although created using direct citation, the topics are very well differentiated from each other textually. 75% of documents with sufficient text can be accurately assigned to the cluster in which they reside using the BM25 similarity measure based on title and abstract. This high level of topic differentiation is extremely important because it allows us to accurately link new documents to individual topics. We take advantage of this in the following way.

The original model only contained publications through 2012. In keeping with our need for stability, and to meet the needs of stakeholders to have an up-to-date model, we decided to add new data to the model rather than creating new models with each annual data update. To update the model, Scopus data from 2013-2015 were added to the existing topics using references and textual information. 7,319,832 source documents and 2,539,272 additional non-source documents that were cited at least twice were assigned to topics using this information. The model through 2015 thus includes a total of 58,257,919 documents, of which 31,935,676 are source documents. For most documents, the topics corresponding to their references were identified, and each document was assigned to the topic containing the largest number of its references. There were a few documents with very few references, but with very long abstracts. These were assigned to the topic with which they had the largest BM25 text similarity. The same process can be used to accurately link other types of documents to the model as well, whether they be articles, patents, or grants. This will be described in more detail in Section 5.

The model also exhibits size distributions that are consistent with what we would expect given other complex networks. Figure 2 shows distributions for topic sizes – numbers of documents and numbers of authors – for 2010 as a function of rank order. When fractionally assigning authors to topics, the largest topic in 2010 had about 3500 authors. 9,400 topics had 100 or more authors, while 33,800 topics had less than 10 authors. In terms of source documents, the largest topic had 940 papers. 4,000 topics had 100 or more documents, while 42,600 topics had fewer than 10 documents in 2010. Both curves are roughly linear on a log-log scale at the upper end of the distribution, and then tail off at the lower end. The fact that some topics are very large, and that others are very small, correlates with Derek de Solla Price's notions of Big Science and Little Science (1963). Big science refers to major investments in infrastructure and the corresponding creation of research communities involving potentially hundreds of researchers. Big Science communities are especially common in biomedical research and high energy physics, and are occasionally found in other fields as well. Significant funding for infrastructure is a necessary pre-requisite for Big Science. Little Science can survive on the existing infrastructure and requires much less external funding.

## 4 Topic Prominence

As mentioned above, we already have an indicator to rank topics by emergence potential. It is also simple to count articles, authors, citations, etc., and to rank topics by those quantities. However, these measures simply reflect size and age. We have been seeking an indicator that would reflect the visibility and/or momentum of topics, and that has the potential to predict whether a topic will grow or decline in the near future, regardless of whether the topic is considered to be emergent or not. This type of indicator could be very useful for stakeholders in their portfolio analysis and planning efforts.

Naming of an indicator is also important because a name can carry unintentional negative connotations. For instance, if we label some topics as 'hot', then we automatically label others as 'cold', and thus suggest they are unimportant and to be avoided. We have chosen to call our new indicator *prominence*, following Grandjean et al. (2011), who found that *prominent* chemicals were well researched, but that there were high priority chemicals (i.e., important from an environmental standpoint) that were not well researched. In our context, *prominence* simply tells us that a topic is visible, and that it has momentum. It tells us whether the broader community of researchers is paying attention to the topic. One key feature of *prominence*, however, is that it cannot be unilaterally equated with importance, innovativeness, newness, or hotness. As with the chemicals in Grandjean et al., a topic can be less prominent but still very important. Importance varies with context, is likely based on stakeholder and societal needs, and is something that we have not yet figured out how to measure.

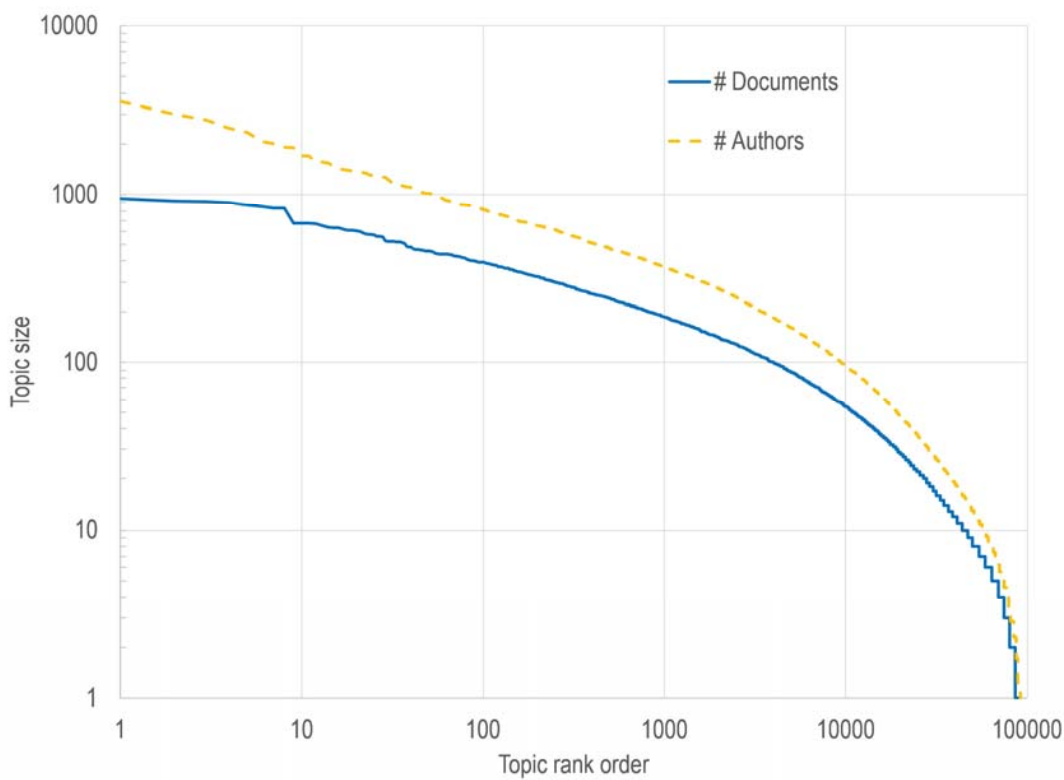

**Figure 2. Topic size distributions for 2010 in terms of documents and authors from the STS model of science.**

In seeking an indicator of prominence, the following variables were considered and were calculated by topic and year *n*:
- Citation counts to articles published in years *n* and *n-1*,
- Scopus Views counts to articles published in years *n* and *n-1*,
- Average CiteScore (Elsevier's new journal-level metric),
- Average number of authors per article,
- Vitality – formerly referred to as thought leadership (Klavans & Boyack, 2008) – essentially a measure of mean reference age,

We also considered including patent reference counts to articles and the fraction of articles authored by industry researchers. However, we ultimately decided against including features related to economic motives in the indicator.

Inclusion of citation counts is an obvious choice, as mentioned above. However, we were also able to consider usage statistics. Elsevier kindly provided us with usage counts by article and month for documents indexed in Scopus, where a usage event was either a click to view an abstract or a click-through to the publisher website to view a full-text copy of the document. Much is known about usage data and how they differ by field (Broad, 1981; Fanelli & Glänzel, 2013; Van Noorden, Maher, & Nuzzo, 2014). However, metrics based on these data have not yet become widespread. Two-year windows were used for the citation and usage counts to focus on recency. Limiting these counts to recently published papers provides a more current depiction of momentum than if a longer time window were used.

We also felt that a journal-level indicator should be considered because it has the potential to capture the social or popular view of science more than would citation counts. We originally worked with three journal metrics – journal impact factor, SNIP, and CiteScore. Upon finding that the three were correlated at the topic level at 0.98 or above, we ultimately chose to use Elsevier's CiteScore metric to maintain consistency with the fact that we use Elsevier's Scopus data. Number of authors was also considered because it has been shown to correlate with impact (Larivière, Gingras, Sugimoto, & Tsou, 2015), and vitality, which is inversely related to mean reference age, was also considered because it favors topics that are quickly building on recent work.

The first four of these candidate variables were log-transformed [log(n+1)] before use because of their extreme skewness. The final variable (i.e., vitality) ranges from zero to one, was not highly skewed, and thus was not transformed. Table 1 shows the correlation between these five variables. Correlations vary, but in general are reasonably high. In particular, L:Views is highly correlated with L:Citations, which is consistent with prior studies (Broad, 1981; Fanelli & Glänzel, 2013). Journal impact (L:CiteScore) is also strongly correlated with L:Citations, but not to the same extent as L:Views. Vitality is the variable that is least correlated with the others.

Although these variables are relatively strongly correlated, that does not automatically mean that they should not all participate in a composite indicator. We thus proceeded with a factor analysis using these five variables to decide which should be included in the composite indicator, the results of which are shown in Table 2. The eigenvalue of the primary factor is 2.33, which is quite high when one is dealing with only five variables. In addition, the factor loadings suggest that the last two variables (authors per paper and vitality) might best be considered independent variables. Given these results, we decided to use only three variables – citations, views and CiteScore – in a composite indicator of topic prominence.

**Table 1. Correlation matrix for candidate variables. "L:" denotes log transform.**

|            | L:Citations | L:Views | L:CiteScore | L:Authors | Vitality |
|------------|-------------|---------|-------------|-----------|----------|
| L:Citations | 1.000 |  |  |  |  |
| L:Views | 0.810 | 1.000 |  |  |  |
| L:CiteScore | 0.533 | 0.483 | 1.000 |  |  |
| L:Authors | 0.395 | 0.395 | 0.509 | 1.000 |  |
| Vitality | 0.313 | 0.288 | 0.290 | 0.425 | 1.000 |

Additional factor analyses were done using only these three variables on data for each of five consecutive years (2008-2012). The resulting factor scores were very similar from year to year, suggesting that relative contributions of the variables are stable and that the indicator can be calculated annually using a single formula. Average normalized factor scores from the three-variable factor analyses are shown in Table 2, and represent the relative weighting of each variable. Citations is the most influential variable and views is almost as important, which suggests that views complements (rather than supplements) the signal from citations. Views counts are important and deserving of more study.

**Table 2. Factor loadings and scoring coefficients used to calculate topic prominence.**

|             | **Factor 1** | **Factor 2** | **Normalized Score** |
|-------------|--------------|--------------|----------------------|
| L:Citations | 0.837 | - 0.244 | 0.495 |
| L:Views | 0.812 | - 0.262 | 0.391 |
| L:CiteScore | 0.653 | 0.154 | 0.114 |
| L:Authors | 0.593 | 0.334 | (not used) |
| Vitality | 0.441 | 0.269 | (not used) |

In equation form, prominence is calculated for each topic $j$ in year $n$ as:

$$P_j = 0.495\,(C_j - mean(C_j))/stdev(C_j) + 0.391\,(V_j - mean(V_j))/stdev(V_j) + 0.114\,(CS_j - mean(CS_j))/stdev(CS_j),$$

where $c_j$ is citation counts to articles in cluster $j$ published in years $n$ and $n-1$, $v_j$ is the Scopus views counts to articles in cluster $j$ published in years $n$ and $n-1$, and $cs_j$ is the average CiteScore for articles in cluster $j$ published in year $n$. These raw values are log-transformed into the values used in the formula as $C_j = ln(c_j + 1)$, $V_j = ln(v_j + 1)$, and $CS_j = ln(cs_j + 1)$. Prominence, as defined here, is a linear combination of citations, views and journal impact for a given topic, where each of these factors is normalized by the topic standard deviation.

## 5 Prediction of Topic-Level Funding

The final step in this development effort was to establish the usefulness of the prominence indicator for the purpose of portfolio analysis. This requires a prediction model – creating an indicator of topic prominence in one time period and correlating that indicator with funding allocations in a following time period. This effort required two steps. First, project-level data were assigned to topics in the model. Once this was done, future funding by topic was correlated with historical prominence values. Additional analyses of the data show that funding per author is strongly related to prominence. Specific examples are provided to illustrate the strengths and weaknesses of the proposed indicator of topic prominence.

**5.1 Assignment of Individual Grants to Topics**

Although funding data are certainly aggregated and reported by high-level field by several institutions (e.g., OECD, U.S. National Science Board), to the best of our knowledge, project-level funding data have rarely, if ever, been analyzed. In fact, relatively little project-level funding data are even publicly available. Aggregated funding data are not suitable for a topic-level analysis. Project-level funding data are required. Fortunately, there is one large repository of project-level funding data that is publicly available. STAR METRICS® is a collaborative effort among many U.S. government agencies to make data and tools publicly available to help assess research impact. For the 2008-2014 time period, it contains project level data (including title, abstract, investigator, funding amount, etc.) for over 364,000 unique grants accounting for $258 billion in funding, or roughly 24% of the U.S. federal R&D budget over that time period. Multi-year grants from the National Institutes of Health (NIH) are listed with separate amounts for each fiscal year, while grants from the National Science Foundation are only listed in the year of their inception with the total grant amount. Notably missing are project-level data from the Departments of Defense and Energy, which have more than double the R&D funding of the agencies represented in STAR METRICS®.

We assigned 314,000 of these grants to topics in the STS model using text from their titles and abstracts. As mentioned above, we used the BM25 textual similarity measure to calculate topic-topic similarities that were used to create the map of Figure 1. We thus already had BM25 word profiles and weights for each topic, and thus calculated BM25 values for each article-topic combination using all articles (around 3.5 million) from two years, 2010 and 2012. Results of this experiment are shown in Figure 3. We found that articles with very high BM25 values were accurately reassigned to their correct topics with over 90% accuracy. Accuracy decreases with decreasing BM25 values. For articles with BM25<40, accuracy drops off very quickly. Overall, nearly 70% of the articles had a BM25$\geq$40, and of these, 75% were accurately reassigned to their correct topic. Given the natural ambiguities in text, we feel that this level of accuracy is quite acceptable for assigning articles to topics using text in the absence of citation information.

Along the same lines, we reasoned that assignment of grants to topics should be similarly accurate because grant titles and abstracts are similar to those of scientific articles in both structure and content. BM25 similarity values were calculated for each grant-topic combination using the grant title and abstract and the stored topic word profiles. In accordance with our findings for articles, only those grants with BM25$\geq$40 were kept for analysis. Table 3 shows the distribution of grants and funding amounts across agencies and years that were used in the analysis. We note that the funding is dominated by NIH, with NSF a distant second. We also note that relatively little (7.2%) of the funding for the Centers for Disease Control and Prevention (CDC) was analyzed because the STAR METRICS® data contain very little text for most CDC grants. Thus, clinical trials are relatively underrepresented. It is interesting to note that the funding amounts were much higher for 2009 and 2010 than for the surrounding years, which had relatively constant funding. This reflects the additional (one-time) investment made in U.S. research and development through the American Recovery and Reinvestment Act of 2009 (ARRA), otherwise known as the economic stimulus package.

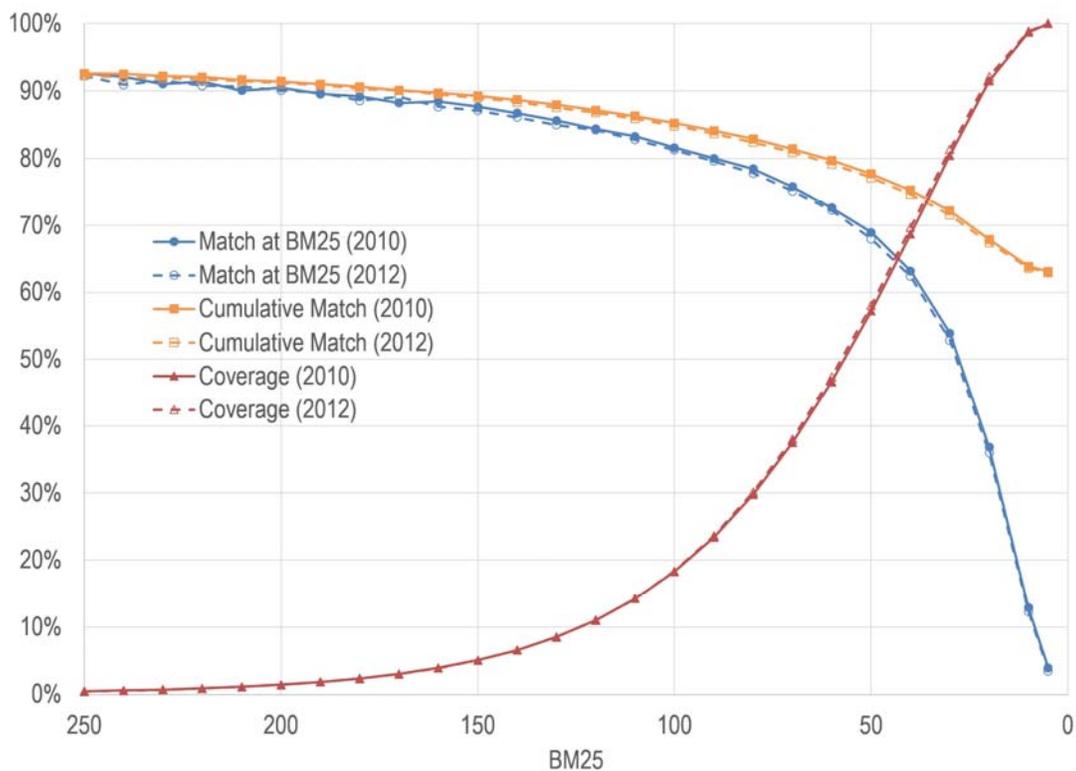

**Figure 3. Accuracy of assigning articles to topics using the BM25 text similarity measure.**

**Table 3. U.S. funding by agency and year that were assigned to topics. Data source: STAR METRICS®. Percent denotes the fraction of funding with sufficient text to be analyzed. Agency results are over all years, and annual results are over all agencies.**

| Agency | # Grants | $ Million | Percent | | Year | # Grants | $ Million | Percent |
|---|---|---|---|---|---|---|---|---|
| NIH | 220,069 | 167,985 | 90.3% | | 2008 | 63,516 | 27,695 | 78.0% |
| NSF | 71,396 | 25,031 | 84.4% | | 2009 | 77,400 | 33,729 | 77.9% |
| CDC | 2,587 | 2,152 | 7.2% | | 2010 | 72,417 | 33,914 | 80.3% |
| NASA | 5,529 | 1,088 | 34.6% | | 2011 | 62,589 | 27,912 | 78.5% |
| NIFA | 6,010 | 2,711 | 60.8% | | 2012 | 61,837 | 26,707 | 77.5% |
| CDMRP | 3,276 | 2,279 | 95.1% | | 2013 | 60,633 | 25,530 | 77.2% |
| Other | 5,228 | 2,187 | 77.3% | | 2014 | 62,928 | 27,947 | 81.9% |
| Total | 314,095 | 203,433 | 78.8% | | Total | 314,095 | 203,433 | 78.8% |

A significant fraction of the funding (particularly for NIH) is tied to large center grants where one can expect the funding to be spread over multiple related topics. Thus, rather than assigning each grant to a single topic, we fractionally assigned the funding for each grant to its top five topic matches, with fractions based on relative BM25 values. We realize this is a somewhat arbitrary choice, but feel it reflects reality better than assigning very large amounts of money to single topics for grants that are obviously multi-topic. One other constraint was placed on the grant-topic assignments. Since these grants are all made by U.S. agencies, very little of the funding is likely to end up in topics where there are no U.S. researchers. Thus, topic choices were restricted to those topics with U.S. researchers in the particular year considered. Once these grant-topic assignments were made, funding was summed by topic and year, and these topic-level data were used in the next step of the analysis.

We note that there are limitations associated with use of these funding data. For instance, these funding data are biased – they represent only one country, and are primarily health-related, ignoring large amounts of funding in defense, energy and intelligence, most of which would go to the natural sciences and engineering. We also know that text-based grant-topic assignments will have many false-positives and false negatives – the rate of inaccurate assignments is at least 25% – and that the inaccurate assignments may themselves have field-level biases. Nevertheless, we demonstrate below that significant insights can be gained from analysis of these data.

**5.2 Funding and Prominence**

Now that we have funding amounts by topic, we can proceed to investigate the correlation between funding and prominence. The structure of this analysis is straightforward. The grant data are split into two time periods (2008-2010 vs. 2011-2013) to reduce the effect of year-to-year variations. Funding amounts were represented in millions of U.S. dollars, and were log transformed [log(n+0.001)] due to their skewness. Topic funding in 2011-2013 was considered the dependent variable, while earlier funding (2008-2010), topic prominence in 2010, vitality, and L:Authors were the independent variables. Correlations between these variables are shown in Table 4.

**Table 4. Correlation matrix for variables considered in the funding prediction analysis.**

|            | L:Fund1113 | L:Fund0810 | Prominence | Vitality | L:Authors |
|------------|------------|------------|------------|----------|-----------|
| L:Fund1113 | 1.000      |            |            |          |           |
| L:Fund0810 | 0.837      | 1.000      |            |          |           |
| Prominence | 0.606      | 0.616      | 1.000      |          |           |
| Vitality   | 0.166      | 0.162      | 0.314      | 1.000    |           |
| L:Authors  | 0.160      | 0.171      | 0.242      | 0.202    | 1.000     |

Not surprisingly, the best predictor of topic funding is historical funding ($R^2 = 0.701$) – funds that were allocated to a topic in the previous time period are a good predictor because many grants span both time periods. The two variables that were excluded from prominence (vitality and author number) have low correlations. We checked this in a regression equation and found that these two independent variables do not contribute to predicting future funding. In contrast, a single-variable prediction model with future funding as a function of historical prominence has an $R^2 = 0.367$. while a two-variable prediction model where historical funding is included increases the $R^2$ to 0.713.

While it is obvious from these data that past funding is the best predictor for future funding. this is not very helpful because most worldwide project-level (and thus topic-level) funding data are not available. A proxy for funding data is needed, and our results suggest that prominence may be a suitable proxy for funding. We are aware, however, that the correlation between funding and prominence may be an effect of topic size. If funding per author is constant with respect to prominence, prominent topics, typically being larger, will receive more money than smaller topics simply because of their size. To check for this possibility, we looked at the funding per author as a function of prominence. We estimated funding per author for each topic by dividing funding by the number of U.S. authors, assuming that the grants reported in STAR METRICS® were being allocated to U.S. authors. Topics were ordered from low to high prominence and then binned so that each bin would have around 25,000 authors. This process generated 31 bins of topics (from the 775,000 U.S. authors) ranked from low to high prominence. Binning topics and summing grants for each bin allowed us to reduce the inherent noise in the data.

The results of these analyses are shown in Figure 4. Starting at the lower left, it is clear that topics with exceptionally low prominence have few U.S. authors as well as very little funding per author. The first bin of 25,000 authors is comprised of the 21,841 topics with the lowest prominence. With the average proposal requiring around 200 hours (worth around $20,000) of investigator time to write (Herbert, Barnett, Clarke, & Graves, 2013; Von Hippel & Von Hippel, 2015), it seems unlikely that a researcher would want to submit a proposal on a topic in this lowest bin, knowing the rate of return is likely to be so low. The data support this reasonable assumption in that only 3.7% of the topics in this low prominence bin received funding.

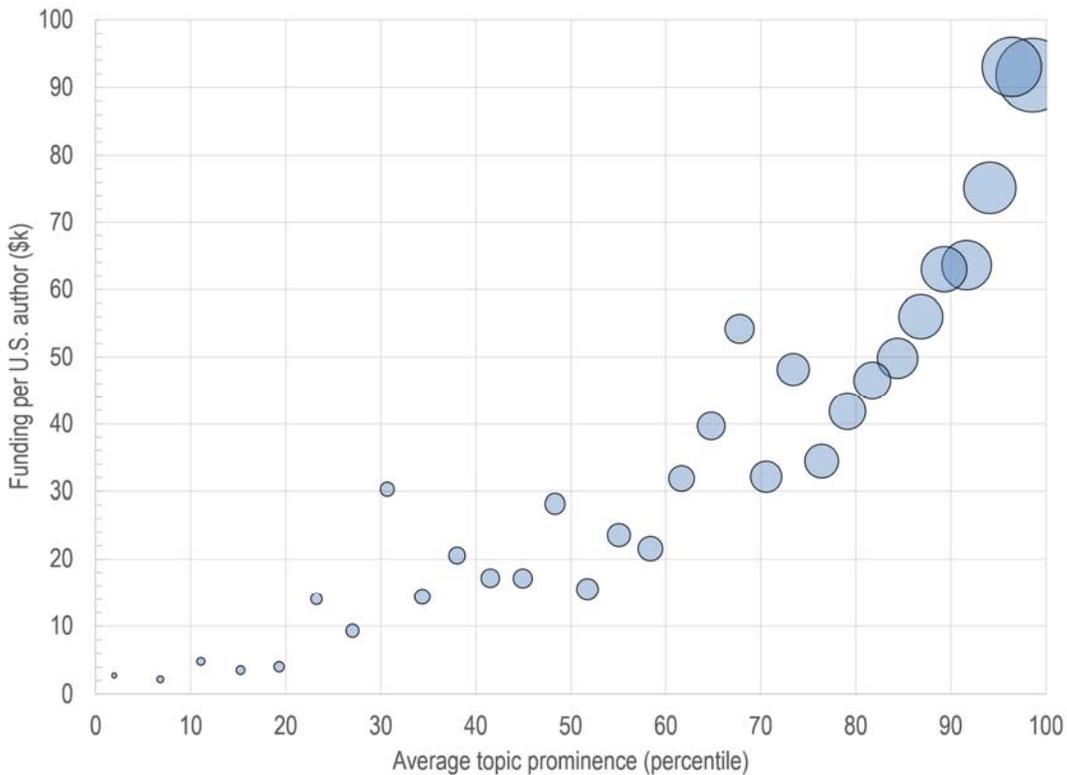

**Figure 4. Funding per U.S. author (2010 only) as a function of topic prominence. Circle size reflects the numbers of authors per topic.**

At the other extreme is the last bin of 25,000 authors. This bin is comprised of the 136 topics with the highest prominence, of which only three topics did not receive funding. These topics had an average of 184 U.S. authors each, who received an average of $91,870 per person (annually 2011-2013) from the funding sources in these data. Clearly, prominent topics should be highly attractive to the individual researcher from a fiscal perspective. It is important to note that this analysis underestimates the actual amount of funding per researcher because only 24% of the U.S. federal R&D budget is considered. Despite limitations in the data and analysis, it seems clear that individual researchers in highly prominent topics receive far more grant funding than those working in topics that are not prominent. Until now there has been no indicator that might alert a researcher about potential funding levels by topic.

### 5.3 Topic Characterization

Although prominence is correlated with funding on the whole, there are topics where there seems to be a mismatch between prominence and STAR METRICS® funding. In this section, we characterize a number of topics to give a practical feel regarding prominence, and to provide evidence of the strengths and weaknesses of the methodology. Table 5 lists ten topics as examples in each of three categories: high prominence high funding (HH), high prominence no funding (HN), and low prominence high funding (LH). The 10 HH topics need little explanation. These have recognizable descriptions, are primarily medical, have many U.S. authors, and have grown significantly from one time period to the next. Topic descriptions are based on characterizations we create for each topic such as the one shown in Figure 5 for the topic on amyloid function in Alzheimer's disease (#205).

**Table 5. Characterization of topics with different prominence and funding levels. Prominence values are listed as percentiles (0 – 100 scale).**

| Topic | #U.S. Author (2010) | Prom 2010 (pctl) | Funding 2010 ($ million) | #Pub (2008-2010) | #Pub (2011-2013) | Description | Discipline |
|---|---|---|---|---|---|---|---|
| **High Prominence – High Funding (HH)** | | | | | | | |
| 2538 | 674.2 | 99.8 | 420.6 | 807 | 2379 | next-generation DNA sequencing | Cell Biology |
| 73 | 364.5 | 99.3 | 305.2 | 1597 | 1960 | T-lymphocytes | Immunology |
| 1544 | 181.7 | 98.3 | 189.4 | 747 | 1297 | orbitofrontal cortex and reward | Neurodeg Diseases |
| 1493 | 223.7 | 99.0 | 180.1 | 742 | 2070 | default mode network (brain) | Brain,Vision,Hearing |
| 2771 | 246.1 | 98.4 | 150.2 | 753 | 1204 | inflammation and obesity | Diabetes |
| 5042 | 209.1 | 95.1 | 143.4 | 396 | 518 | autism phenotype | Psychiatry |
| 236 | 338.7 | 99.1 | 80.1 | 1215 | 1675 | peptide identification in proteomics | Analytical Chemistry |
| 205 | 216.5 | 98.2 | 41.6 | 1053 | 1523 | amyloid function in Alzheimer's | Neurodeg Diseases |
| 2646 | 215.1 | 98.1 | 37.3 | 677 | 945 | solid-state nanopores | Nanochemistry |
| 2877 | 128.3 | 96.4 | 33.0 | 472 | 912 | BPA and endocrine disruption | Environ Chemistry |
| **High Prominence – No Funding (HN)** | | | | | | | |
| 2187 | 0.5 | 95.8 | 0 | 591 | 849 | electrochem degradation in wastewater | Electrochemistry |
| 25 | 1.6 | 96.5 | 0 | 785 | 1532 | corrosion inhibitors (steel) | Materials |
| 135 | 2.7 | 98.2 | 0 | 1118 | 2136 | dye remediation in effluents | Electrochemistry |
| 15 | 4.2 | 98.7 | 0 | 1263 | 1961 | biosorption of heavy metals | Electrochemistry |
| 4003 | 7.8 | 98.8 | 0 | 547 | 878 | dispersive liquid-liquid micro-extraction | Environ Chemistry |
| 566 | 9.2 | 94.7 | 0 | 608 | 753 | properties of olive extracts | Animal Science |
| 4594 | 14.4 | 96.9 | 0 | 540 | 841 | hollow nanoparticles | Electrochemistry |
| 580 | 21.1 | 96.8 | 0 | 1010 | 1847 | phosphors for LEDs | Optical Materials |
| 644 | 21.1 | 95.4 | 0 | 725 | 777 | hydrogen energy storage | Electrochemistry |
| 7 | 74.3 | 99.3 | 0 | 1963 | 2407 | Zn0 nanostructures | Semicond Physics |
| **Low Prominence – High Funding (LH)** | | | | | | | |
| 83249 | 0.1 | 0.3 | 11.9 | 6 | 10 | loose topic - clinical investigation centers | Patient Care |
| 25667 | 2.2 | 2.2 | 24.8 | 6 | 13 | academic medical centers, enthusiasm | Patient Care |
| 54378 | 2.3 | 12.8 | 89.3 | 20 | 20 | forestry education | Agricultural Policy |
| 38569 | 2.7 | 14.2 | 13.7 | 27 | 29 | agroecology, sustainability | Agricultural Policy |
| 54158 | 5.0 | 4.6 | 12.0 | 18 | 60 | loose topic – DMZ, networks, protocols | Computing |
| 33105 | 10.9 | 6.1 | 29.6 | 42 | 60 | role of nursing in clinical trials | Patient Care |
| 18741 | 30.1 | 3.5 | 23.0 | 71 | 137 | capstone projects, engineering | Learning |
| 33702 | 30.4 | 11.2 | 18.3 | 51 | 77 | extension programs and learning | Agricultural Policy |
| 38645 | 31.1 | 8.6 | 10.9 | 72 | 92 | systems engineering competency training | Management |
| 26483 | 39.8 | 9.5 | 44.2 | 99 | 176 | civil engineering program criteria | Learning |

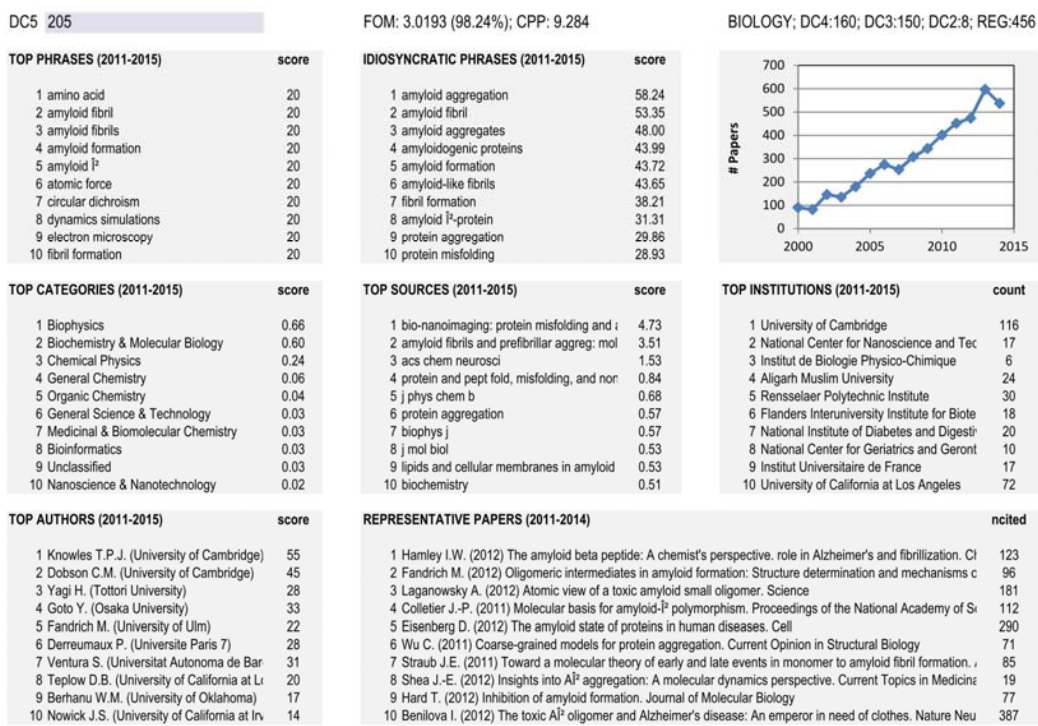

**Figure 5. Characterization of topic #205 on amyloid function in Alzheimer's. Lists of the top phrases, journal categories, sources, papers, etc. associated with the topic can be viewed by the interested user to understand what the topic is about. The chart with number of papers by year shows the rate of growth or decline in a topic over time.**

The second set of topics – the HN topics – is more interesting. The lack of funding does not appear to be associated with assignment errors. Rather, these are all large topics (worldwide publication numbers) with few U.S. authors, and that in some cases reflect regional issues. Why isn't the U.S. involved in these research areas? From this limited sample, it seems that these are research activities that are simply not priorities of the NIH or NSF. For instance, the first five HN topics are remediation efforts related to industries with high levels of pollution. One could argue that the lack of U.S. funding is due to the fact that Environmental Protection Agency (EPA) projects are not included in STAR METRICS® data. However, the lack of funding is reflected in the lack of U.S. authors participating in these very large research topics. If there had been 40 or 50 U.S. authors in any of these topics one could argue that this was due to limitations in the coverage of the funding data. With less than 8 U.S. authors per topic, the stronger argument is that these topics are not priorities in the U.S., and that its authors don't choose to work on them.

A similar argument can be made for the sixth HN topic which is related to olive oil. The U.S. is not a major producer of olive oil and U.S. government agencies are correspondingly not funding this area. The final four HN topics are all related to materials or nanotechnology. These are large topics whose major contributors are universities in the Asia-Pacific region, with Chinese universities taking the top positions. While there are U.S. authors in these topics, it is very possible that funding for U.S. work in these topics comes from industry or defense sources. Other nations have, however, made research in these topics more of a national priority. In general, all ten HN outliers from Table 5 seem to be accurate representations of national research policies.

The ten examples of LH topics tell a completely different story. These data suggest that small topics with low prominence that are characterized with ambiguous terms are inaccurately linked to grants. Specifically, the very small topics dealing with clinical trials and medical centers (#83249, #25667, #33105), education (#54378, #18741, #26483), and agricultural policy (#38569, #33702) are likely improperly linked to grants because of language ambiguity. For example, there are only six articles that were published in topic #83249 between 2008 and 2010. None of these articles was published by a US author. However, 'clinical trial' and 'regulatory compliance' are two of the phrases that were common to this very small set of documents. These phrases are also relatively common in grant documents, and have seemingly resulted in the assignment of $12 million dollars of U.S. grant money (2011-2013) to this topic. We also note that U.S. grant applications have for many years required the addressing of broader impacts, of which education is a primary constituent. Thus, it is not surprising that topics in education are being assigned more grant money using our methodology than perhaps is accurate.

The LH set also includes small, low prominence topics that contain ambiguous phrases. Perhaps the best example is topic #54158, which is a very loose collection of documents about DMZ networks, protocols, and security. There are other, much more prominent and focused topics on networks and security. In summary, topics that have high prominence are typically large, and seem to be accurately linked to grants because there is more text to work with. The lack of funding of highly prominent topics in the HN set reflects a reasonable characterization of U.S. government research policy. There seem to be good reasons why the U.S. is not funding some highly prominent research. However, in the case of low prominence topics with much higher than expected levels of funding, we suspect that these reflect inaccurate linking of grants to topics due to textual ambiguity. In these cases, we suspect that the prominence metric is relatively accurate, and that the grant assignments are not. This needs more detailed exploration. Methods for properly analyzing and characterizing (apparently) low prominence topics are needed.

# 6 Summary and Implications

In this article, we have addressed two gaps in the research planning system. The first is associated with the current supply of science, and the need for a detailed classification of science at the topic level to enable more targeted decision making by stakeholders in the science system. We have presented a method for partitioning a large citation database into topic-level structures, and have justified our choice of creating a model with approximately 100,000 topics. Our current model, consisting of 91,726 topics that contain over 58 million documents has been detailed, along with examples of how it can be used in practice.

The second gap is associated with the future demand for science. We believe it is fair to say that bibliometrics has become overly focused on the issue of fair evaluation to the detriment of the practical need to plan for the future. We have presented a new indicator of topic prominence and have shown that this indicator may have predictive value. Prominence, by itself, explains over one-third of the variance in future funding by topic, and topics with high prominence receive far more funding per author than topics with low prominence. These results suggest that prominence is ultimately an indicator of demand for science as represented by where funders choose to allocate funds. We note that this does not necessarily represent the breadth of demand that is associated with societal goals. We also present examples of topics where prominence and funding are not well correlated, and provide reasonable explanations for the differences.

There are a number of methodological strengths and limitations in this study. On the supply side, major strengths include the use of a comprehensive database (Scopus) that can be assumed to cover the majority of the research topic space, and our decades of experience in generating document clusters that correspond to research communities. A corresponding weakness is the lack of coverage in specific areas such as agriculture. Even when using one of the databases with the broadest coverage, there are topics that may be missed using our global approach. For cases such as this where detailed knowledge of the science supply is needed in an area where the coverage of WoS or Scopus is less than desirable, use of a specialized database (such as CABI) is an attractive alternative. Another approach would be to add the non-duplicative content from a specialized database into a global model using textual similarity, and to create new topics from those papers that cannot be strongly linked into the global model.

On the demand side, the major strength is the procedure that assigns hundreds of thousands of grants to topics. We have assumed that the funding agencies made their decisions based on the individual merit of each proposal – that is they imposed local criteria about the value of this research for a variety of stakeholders. Therefore, the funding allocations across all topics represent an overall consensus of demand from the perspective of these funders. The fact that topic prominence can predict these future research allocations is very useful. Nevertheless, we must not lose sight of the fact that prominence represents overall demand and visibility, and does not necessarily reflect importance. Indeed, science may generate its own priorities rather than basing demand on social needs (Evans, et al., 2014). Much more must be done to understand the various dimensions of importance, especially to the public, and how they can be measured.

Two of the methodological weaknesses in this study were the inaccurate assignment of some grants to topics, and the lack of comprehensiveness of the funding data. We showed that articles can be assigned to topics with 75% accuracy, and assumed that the same would hold for grants. However, we also showed examples where grants were improperly assigned to topics of low prominence because those topics were associated with common text (e.g., "clinical trials") in a general way. These inaccurate assignments may be affecting the correlation between funding and prominence. However, at present we have no way of knowing if this effect would be positive or negative. The assignment of grants to topics needs to be improved, and further study in this area is needed.

Regarding funding data, although the amount of data used was substantial, it is still only a fraction of all funding data, and is strongly biased to health-related funding in the U.S. Funding for topics associated with energy and defense are not included. Once again, this could affect the correlations between prominence and funding, and further data and study are needed to strengthen our conclusions. However, even though more data would clarify the situation further, we do not expect the addition of defense funding data to significantly change the results because most defense funded research does not end up in the open scholarly literature.

One potential weakness in the results is that, while there is clear correlation between prominence and funding, it is much less clear that prominence affects period-to-period changes in funding levels. Historical trends dominate – the best predictor of future funding is past funding – and the increase in predictive value due to prominence (from $R^2 = 0.701$ to $0.713$), while statistically significant, is marginal. Further work is needed to determine if prominence can predict changes in funding. Despite this weakness, the fact that funding per author increases significantly with an increase in topic prominence suggests that prominence is a useful indicator.

There are also conceptual weaknesses. Most importantly, publication impact (in the many forms used to create the indicator of topic prominence) is not an essential ingredient for measuring the importance of research. Importance is relative to the strategic goals of different institutions and to societal values. For example, for an institution interested in curing cancer, importance and societal impact should be more about whether cancers are being cured that it is about cancer articles being highly cited or cancer researchers receiving an award. For the institution developing artificial intelligence technology, importance and impact should be about subsequent innovations and corresponding economic benefits, not whether an article is highly downloaded. For institutions doing research on challenges such as global warming or terrorism, true impact can only be measured in terms of benefits to society, and not by a journal impact score.

Despite the challenges associated with quantifying impact, we feel the use of topics and prominence has potential benefits to stakeholders in the science system in terms of research planning. For the academic community involved with bibliometric analysis, we acknowledge the tremendous efforts aimed at research evaluation. We correspondingly point out that relatively little work has been devoted to the need for research planning. For practitioners, illustrating how bibliometric tools can predict where future funds are likely to be reallocated is extremely important for survival in a highly competitive funding environment. A detailed model of science coupled with an indicator that predicts funding can be used by funders as an independent measure of potential trends. It can be used by research administrators to identify research topics and researchers that are likely to be associated with prominent topic. This knowledge has the potential to raise the institutional profile over time. Researchers can benefit by knowing which topics are likely to have increased funding, and can tune their proposals to those topics to secure greater funding and advance their careers.

We note, however, that these results should be used with some caution. It is well known that the potential exists for misuse of, or for blind acceptance of, metrics (Hicks, Wouters, Waltman, de Rijcke, & Rafols, 2015). A decrease in the diversity of science (an undesirable outcome) is possible if all, or even a significant fraction, of researchers were to blindly employ prominence for decision making. We think, however, that only a small fraction of researchers will use prominence as the primary source to guide their actions. The researchers we have dealt with over the years have a keen sense of what they consider to be important. These are people who, when faced with a choice between what they consider to be important and what they think will bring in funding, will either choose importance or will find a way to bridge the two. This latter behavior is what we want to encourage.

Also, regarding whether a metric will find wide adoption in practice, we have another example to share that, although from sports, seems relevant. The book "Moneyball" (Lewis, 2003) tells the story of how advanced metrics were introduced in the management of a baseball franchise, and how this team with very limited resources used metrics to have success beyond what was expected given its budget. Given its sustained success, one might think that all teams would adopt similar tactics. However, this has not happened. Twenty years later, although all teams look at such metrics, there are only few teams that base their decisions of who to draft and who to trade primarily on these metrics. Most teams consult metrics, but base their final decisions on the observations of their scouts. In other words, most baseball teams still base their actions on human inputs rather than metrics.

In an analogous manner, we suggest that most researchers will not (and should not) base their decisions solely on a metric such as prominence, but that they will use their knowledge and sense to identify those things that are important, and weigh both human and metric inputs together when making decisions. We also suggest that research administrators will not base their decisions for promotion and tenure on a metric such as prominence, but will rather consider human and metric inputs.

We are hopeful that knowledge of topics and prominence will lead to a higher rate of advance in science overall. We also anticipate development of additional indicators in the future that will address true impact rather than citation counts, and look forward to applying such measures to topic-level structures in science in the hope that they will further enhance decision making.

**Competing Interests**

The work reported here is the result of research conducted over several years, and which was funded internally by SciTech Strategies, Inc. Recently, however, Elsevier BV and SciTech have entered into a contractual arrangement. As a result, topics (based on updated data) and the prominence metric are now available in Elsevier's SciVal product, and SciTech will support Elsevier's efforts related to "Topics of Prominence" for a period of time going forward.